\documentclass{article}
\usepackage[T1]{fontenc}
\usepackage{wrapfig}
\usepackage[portuges,english]{babel}
\usepackage{amssymb,latexsym,amsmath,color,mathrsfs,ifsym,graphics,stmaryrd} 
\usepackage[colorlinks,linkcolor=blue,urlcolor=blue,citecolor=black,
plainpages=false,pdfpagelabels,breaklinks]{hyperref}

\title{{\sc Against `Particle Metaphysics' and `Collapses'\\ within the Definition of Quantum Entanglement}}

\author{{\sc C. de Ronde}$^{1,2,3}$ and {\sc C. Massri}$^{4,5}$}
\date{}

\usepackage[margin=2.1cm]{geometry}

\begin{document}

\bibliographystyle{plain}
\maketitle

\begin{center}
\begin{small}
1. Philosophy Institute Dr. A. Korn, University of Buenos Aires - CONICET\\
2. Center Leo Apostel for Interdisciplinary Studies\\Foundations of the Exact Sciences - Vrije Universiteit Brussel\\
3. Institute of Engineering - National University Arturo Jauretche.\\
4. Institute of Mathematical Investigations Luis A. Santal\'o, UBA - CONICET\\
5. University CAECE
\end{small}
\end{center}

\bigskip

\begin{abstract}
\noindent In this paper we argue against the orthodox definition of {\it quantum entanglement} which has been explicitly grounded on several ``common sense'' (metaphysical) presuppositions and presents today serious formal and conceptual drawbacks. This interpretation which some researchers in the field call ``minimal'', has ended up creating a narrative according to which QM talks about ``small particles'' represented by {\it pure states} (in general, superpositions) which ---each time someone attempts to observe what is going on--- suddenly ``collapse'' to a single measurement outcome. After discussing the consequences of applying `particle metaphysics' within the definition of entanglement we turn our attention to two recent approaches which might offer an interesting way out of this metaphysical conundrum. Both approaches concentrate their efforts in going beyond the notion of `system'. While the first, called device-independent approach, proposes an operational anti-metaphysical scheme in which language plays an essential role; the second approach, takes an essentially metaphysical path which attempts to present a new non-classical representation grounded on {\it intensive relations} which, in turn, imposes the need to reconsider the definition and meaning of quantum entanglement.  
\end{abstract}
\begin{small} 

\end{small}

\newtheorem{theo}{Theorem}[section]
\newtheorem{definition}[theo]{Definition}
\newtheorem{lem}[theo]{Lemma}
\newtheorem{met}[theo]{Method}
\newtheorem{prop}[theo]{Proposition}
\newtheorem{coro}[theo]{Corollary}
\newtheorem{exam}[theo]{Example}
\newtheorem{rema}[theo]{Remark}{\hspace*{4mm}}
\newtheorem{example}[theo]{Example}
\newcommand{\proof}{\noindent {\em Proof:\/}{\hspace*{4mm}}}
\newcommand{\qed}{\hfill$\Box$}
\newcommand{\ninv}{\mathord{\sim}} 
\newtheorem{postulate}[theo]{Postulate}

\bigskip

\section*{Introduction}

The notion of {\it entanglement} plays today the most central role in what might be regarded to be the origin of a new groundbreaking technological era. In the specialized literature, this large field of research falls under the big umbrella of what is called {\it quantum information processing}. This term covers different outstanding non-classical technical developments such as, for example, quantum teleportation, quantum computation and quantum cryptography. Without exception, all these new technologies are founded on the notions of {\it quantum superposition} and {\it entanglement}. Even though today, technicians, computer scientists, cryptographers and engineers are rapidly advancing in the creation of new technological devices and algorithms, it is important to stress the fact that the notion of entanglement remained almost unnoticed for half a century since its coming into life in 1935 when Albert Einstein, Boris Podolsky and Nathan Rosen discussed what would later become known as the famous EPR {\it Gedankenexperiment}. In that same year, Erwin Schr\"odinger in a series of papers gave its name to the new born concept. He discussed in depth what he called {\it entanglement} ({\it Verschr\"rankung} in German) and showed through the now famous `cat paradox' how quantum correlations rapidly expanded into the classical domain. This new concept was not seriously considered by the community of physicists which thought that Bohr had solved everything through the principle of complementarity. But after half a century, with the new technical possibilities, what Einstein, Podolsky, Rosen and Schr\"odinger had critically imagined, was now experimentally testable in the lab. And the results of experiments confirmed, against their classical prejudices, the predictions of the theory of quanta. Suddenly, entanglement became one of the key concepts of QM, not only in foundational territory but also in technological ones. However, regardless of its growing influence, still today, the concept of entanglement presents deep mathematical and conceptual difficulties. In this paper we will argue that the reason behind these difficulties are related to the inadequate ---already pointed out by both Einstein and Schr\"odinger--- addition of `particle metaphysics' and `invisible collapses' within the orthodox axiomatic formulation of the theory.  

The paper is organized as follows. In section 1, we discuss the original definition of entanglement as presented in 1935 by Einstein, Podolsky, Rosen in \cite{EPR} and Schr\"odinger in \cite{Schr35a}. In section 2, we present the contemporary definition of quantum entanglement grounded on the notion of {\it separability}. Section 3 discusses if the addition of the metaphysical picture of particles is really adequate for understanding the orthodox formalism of the theory of quanta. In section 4 we analyze the way in which device-independent approaches might offer a way out of this metaphysical conundrum by escaping the reference to `systems' right from the start. In section 5, we turn our attention to the role played by metaphysics within physics, and in section 6, we analyze the way in which a new (non-classical) metaphysical account of QM imposes the need to redefine the notion of quantum entanglement right from the start. Section 7 presents the conclusions of the paper.

\section{The Critical Origin of Quantum Entanglement}

Critical thought is, above all, the possibility of analysis of the foundation of thought itself. The analysis of the conditions under which thinking becomes possible. By digging deeply into the basic components of thinking, one is able to understand the preconditions and presuppositions which support the architecture of argumentation itself. In this work we attempt to provide a critical analysis of the definition of quantum entanglement which has as one its main cornerstones the EPR {\it Gedankenexperiment} presented by Einstein, Podolsky and Rosen in \cite{EPR}. Many analysis of the mentioned thought-experiment have been already provided within the foundational literature ---between many others--- by Diederik Aerts \cite{Aerts84a, Aerts84b} and Don Howard \cite{Howard85}. In the following, we attempt to extend this analysis paying special attention to the notion of {\it separability} and to the famous definition of {\it element of physical reality} in order to approach ---later on--- the presuppositions involved within the definition of quantum entanglement itself.   

Even though Albert Einstein was certainly a revolutionary in many aspects of his research, he was also a classicist when considering the preconditions of physical theories themselves. His dream to create a unified field theory was grounded in his belief that physical theories, above all, must always discuss in terms of specific situations happening within space and time. In this respect, the influence of transcendental philosophy in Einstein's thought cannot be underestimated \cite{Howard94}. That space and time are the {\it forms of intuition} that allow us to discuss about objects of experience was one of the most basic {\it a priori} dictums of Kantian metaphysics, difficult to escape even for one of the main creators of relativity theory. In a letter to Max Born dated 5 April, 1948, Einstein writes:
\begin{quotation}
\noindent {\small ``If one asks what, irrespective of quantum mechanics, is characteristic of the world of ideas of physics, one is first stuck by the following: the concepts of physics relate to a real outside world, that is, ideas are established relating to things such as bodies, fields, etc., which claim a `real existence' that is independent of the perceiving subject ---ideas which, on the other hand, have been brought into as secure a relationship as possible with the sense-data. It is further characteristic of these physical objects that they are thought of as arranged in a space-time continuum. An essential aspect of this arrangement of things in physics is that they lay claim, at a certain time, to an existence independent of one another, provided these objects `are situated in different parts of space'. Unless one makes this kind of assumption about the independence of the existence (the `being-thus') of objects which are far apart from one another in space ---which stems in the first place in everyday thinking--- physical thinking in the familiar sense would not be possible. It is also hard to see any way of formulating and testing the laws of physics unless one makes a clear distinction of this kind.'' \cite[p. 170]{Born71}}
\end{quotation}

\noindent This precondition regarding objects situated in different parts of space can be expressed, following Howard \cite[p. 226]{Howard89}, as a principle of spatio-temporal separability:

\smallskip
\smallskip

\noindent {\it {\bf Separability Principle:} The contents of any two regions of space separated by a non-vanishing spatio-temporal interval constitute separable physical systems, in the sense that (1) each possesses its own, distinct physical state, and (2) the joint state of the two systems is wholly determined by these separated states.}

\smallskip
\smallskip

\noindent In other words, the presence of a non-vanishing spatio-temporal interval is a sufficient condition for the individuation of physical systems and their associated states. Everything must ``live'' within space-time; and consequently, the characterization of every system should be discussed in terms of {\it yes-no questions} about physical properties. But, contrary to many, Einstein knew very well the difference between a conceptual presupposition of thought and the conditions implied by mathematical formalisms. In this respect, he also understood that his principle of separability was {\it only for him} a necessary metaphysical condition for doing physics. More importantly, he was aware of the fact there was no logical inconsistency in dropping the separability principle in the context of QM. At the end of the same letter to Born he points out the following:
\begin{quotation}
\noindent {\small ``There seems to me no doubt that those physicists who regard the descriptive methods of quantum mechanics as definite in principle would react to this line of thought in the following way: they would drop the requirement for the independent existence of the physical reality present in different parts of space; they would be justified in pointing out that the quantum theory nowhere makes explicit use of this requirement.'' \cite[p. 172]{Born71}}
\end{quotation}

\noindent This famous passage shows that Einstein was completely aware of the fact that QM is not necessarily committed to the metaphysical presupposition of space-time separability. But let us now turn to the kernel of the EPR argument, namely, their introduction of a ``reality criterion'' which would stipulate a sufficient condition for considering an element of physical reality:\footnote{We are thankful to Prof. Don Howard for pointing us the specificity of the reality criterion.}

\smallskip
\smallskip

\noindent {\it {\bf Element of Physical Reality:} If, without in any way disturbing a system, we can predict with certainty (i.e., with probability equal to unity) the value of a physical quantity, then there exists an element of reality corresponding to that quantity.}

\smallskip
\smallskip

\noindent This definition introduced a co-relation between, on the one hand, a {\it certain prediction}, and on the other, the value of a physical quantity (or property) of a system. Certainty is then understood as probability equal to unity. Notice that this remark is crucial in order to filter the predictions provided by QM. Only those related to probability equal to one, $p = 1$, can be considered to be related to physical reality. This means, implicitly, that the rest of the quantum mechanical probabilistic predictions which are not equal to one ---namely, those which pertain the interval between 0 and 1---, $p \in (0,1)$, are simply not considered. Given a quantum state, $\Psi$, there is only one meaningful operational statement (or property) that can be predicted with certainty. This has lead to the conclusion that only one property can be regarded as being {\it actual} (or real).\footnote{This idea has been strongly endorsed by Dennis Dieks within his neo-Bohrian modal interpretation.} While the rest of quantum properties are considered as being {\it indeterminate}. The important point is that the ``non-certain'' predictions are not directly related to physical reality. Unlike real actual properties, indeterminate properties are considered as being only ``possible'' or ``potential'' properties; i.e., properties that might become actual in a future instant of time (see for a detailed analysis \cite{Sudbery16}). Until these properties are not actualized they remain in a sort of limbo. As Heisenberg \cite[p. 42]{Heis58} explains, such properties stand ``in the middle between the idea of an event and the actual event, a strange kind of physical reality just in the middle between possibility and reality.'' The filtering of indeterminate properties ---something which, at least from an operational perspective, seems completely unjustified---, is directly related to the actualist spatio-temporal (metaphysical) understanding of physical reality which Einstein so willingly wanted to retain. As he made the point \cite{Howard17}:  ``that which we conceive as existing (`actual') should somehow be localized in time and space. That is, the real in one part of space, $A$, should (in theory) somehow `exist' independently of that which is thought of as real in another part of space, $B$.''\footnote{As we discussed in detail in \cite{deRondeMassri18}, it is not difficult to see that this actualist understanding of existence is grounded in the classical representation of physics provided in terms of an {\it actual state of affairs} and {\it binary valuations}.} 

But, as noticed by Bohr himself in his famous reply to EPR \cite{Bohr35}, it is the first part of the definition which introduces a serious ``ambiguity''. Indeed, the previous specification, {\it ``If, without in any way disturbing a system,''} refers explicitly to the possibility of measuring the system in question. It thus involves an improper scrambling between ontology and epistemology, between physical reality and measurement. A scrambling ---let us stress---, completely foreign to all classical physics. This scrambling, might be regarded as one between the many ``quantum omelettes'' created during the early debates of the founding fathers \cite[p. 381]{Jaynes}. However, it might be also interesting to notice that the EPR criteria goes against one of Einstein's most interesting characterizations of physical theories \cite[p. 175]{Dieks88a} : ``[...] it is the purpose of theoretical physics to achieve understanding of physical reality which exists independently of the observer, and for which the distinction between `direct observable' and `not directly observable' has no ontological significance''. This is of course, even though ``the only decisive factor for the question whether or not to accept a particular physical theory is its empirical success.'' The physical representation of a physical theory is always {\it prior} to the possibility of epistemic inquiry of which `measurement' is obviously one of its main ingredients. As he also remarked to a young Heisenberg: it is only the theory which decides what can be observed, and not the observations which determine the theory.\footnote{This might be one of the main reasons why Einstein did not like the reality criteria presented in the EPR paper. For a detailed analysis of the disagreements of Einstein with the EPR paper see: \cite{Howard85}.} 

The collapse of the quantum wave function was added to the axiomatic formulation of the theory in order to account for single measurement outcomes ---evading the theoretical reference to quantum superpositions (see for a detailed analysis: \cite{deRonde18}). Since, in a positivist fashion, it was assumed that physical theories should be able to describe observations, the famous {\it projection postulate} became a necessary condition in order to secure the empirical content of the theory. This was explicitly considered by Paul Dirac and John von Neumann in their famous books at the beginning of the 1930s. As von Neunmann's \cite[p. 214]{VN} made the point: ``Therefore, if the system is initially found in a state in which the values of $\mathcal{R}$ cannot be predicted with certainty, then this state is transformed by a measurement $M$ of $\mathcal{R}$ into another state: namely, into one in which the value of $\mathcal{R}$ is uniquely determined. Moreover, the new state, in which $M$ places the system, depends not only on the arrangement of $M$, but also on the result of $M$ (which could not be predicted causally in the original state) ---because the value of $\mathcal{R}$ in the new state must actually be equal to this $M$-result.'' Dirac \cite[p. 36]{Dirac74} justified the quantum jump in his own manner: ``When we measure a real dynamical variable $\xi$, the disturbance involved in the act of measurement causes a jump in the state of the dynamical system. From physical continuity, if we make a second measurement of the same dynamical variable $\xi$ immediately after the first, the result of the second measurement must be the same as that of the first.'' 

Both Einstein and Schr\"odinger were strongly against the existence of such ``invisible collapses''. Their analysis attempted to show the inconsistencies of assuming such subjectively induced reality. In particular, the EPR {\it Gedeankenexperiment} makes explicit use of the collapse in order to show the strange non-local influence that appears when measuring one of the particles on the other distant (entangled) partner. Indeed, if one accepts the orthodox interpretation of QM according to which the measurement of a quantum superposition induces a ``collapse'' to only one of its terms, Einstein, Podolsky and Rosen then show that there seems to exist a super-luminous transfer of information (or interaction) from one particle to the other distant partner. Once the entangled particles are separated, all their properties still remain {\it indeterminate}. But, the moment we perform a measurement of an observable in one of the particles we also find out instantaneously what is the value of the distant partner ---in case we would choose to measure the same observable. Every time we measure an observable in one of the particles, the other particle ---as predicted by QM--- will be found to possess a strictly correlated value.\footnote{Let us remark that observability is used in this case a sufficient condition to define reality itself. There is involved here a two sided definition of what accounts for physical reality, either in terms of computing the certainty of an outcome (= real) or by observing an outcome which was uncertain but became actual (= real).} Thus, the (real) ``collapse'' of one of the particles also produces the (real) ``collapse'' of the other distant entangled particle. Einstein was of course clearly mortified by this seemingly non-local ``quantum effect'' which he ironically called {\it spukhafte Fernwirkung}, translated later as ``spooky action at a distance''.

That same year, continuing the EPR's critical reflections, Erwin Schr\"odinger introduced in an explicit manner the notion of notion of {\it entanglement}. 
\begin{quotation}
\noindent {\small ``When two systems, of which we know the states by their respective representatives, enter into temporary physical interaction due to known forces between them, and when after a time of mutual influence the systems separate again, then they can no longer be described in the same way as before, viz. by endowing each of them with a representative of its own. I would not call that one but rather the characteristic trait of quantum mechanics, the one that enforces its entire departure from classical lines of thought. By the interaction the two representatives
(or $\psi$-functions) have become entangled.'' \cite[p. 555]{Schr35b}}
\end{quotation}
\noindent Making explicit reference to the EPR paper, Schr\"odinger remarks that: 
\begin{quotation}
\noindent {\small ``Attention has recently \cite{EPR} been called to the obvious but very disconcerting fact that even though we restrict the disentangling measurements to one system, the representative obtained for the other system is by no means independent of the particular choice of observations which we select for that purpose and which by the way are entirely arbitrary. It is rather discomforting that the theory should allow a system to be steered or piloted into one or the other type of state at the experimenter's mercy in spite of his having no access to it.'' \cite[pp. 555-556]{Schr35b}}
\end{quotation}
Following the reality criteria proposed in the EPR paper, Schr\"odinger also assumed ---implicitly--- that {\it maximal knowledge} had to be understood as {\it certain knowledge}; i.e., as knowledge involving probability equal to unity. As he critically remarks, the astonishing aspect of QM is that when two systems get entangled through a known interaction, the knowledge we have of the parts might anyhow decrease. 
\begin{quotation}
\noindent {\small ``If two separated bodies, each by itself known maximally, enter a situation in which they influence each other, and separate again, then there occurs regularly that which I have just called {\it entanglement} of our knowledge of the two bodies. The combined expectation-catalog consists initially of a logical sum of the individual catalogs; during the process it develops causally in accord with known law (there is no question of measurement here). The knowledge remains maximal, but at the end, if the two bodies have again separated, it is not again split into a logical sum of knowledges about the individual bodies. What still remains {\it of that} may have become less than maximal, even very strongly so.---One notes the great difference over against the classical model theory, where of course from known initial states and with known interaction the individual states would be exactly known."\cite[p. 161]{Schr35a}}
\end{quotation}
\noindent It is the projection postulate interpreted as a ``real collapse'' of the quantum wave function which ends up scrambling ---just like in the case of the measurement process--- the objective theoretical representation provided by the mathematical formalism and the subjective observation of a particular `click' in the lab. The entanglement of systems and outcomes within the same representation determines then the scrambling of the objective knowledge, related to the theory, and the subjective knowledge, related to the purely epistemic process of measurement. 

Einstein's criticism against this ``spooky action'', was also a criticism against the addition of a subjectively produced ``collapse''. In this respect, Einstein is quoted by Everett \cite[p. 7]{OsnaghiFreitasFreire09} to have said that he ``could not believe that  a mouse could bring about drastic changes in the universe simply by looking at it''. Schr\"odinger would also make fun of the existence of such induced collapses:
\begin{quotation}
\noindent {\small``But jokes apart, I shall not waste the time by tritely ridiculing the attitude that the state-vector (or wave function) undergoes an abrupt change, when `I' choose to inspect a registering tape. (Another person does not inspect it, hence for him no change occurs.) The orthodox school wards off such insulting smiles by calling us to order: would we at last take notice of the fact that according to them the wave function does not indicate the state of the physical object but its relation to the subject; this relation depends on the knowledge the subject has acquired, which may differ for different subjects, and so must the wave function.'' \cite[p. 9]{OsnaghiFreitasFreire09}} \end{quotation}

\section{The Contemporary Definition of Quantum Entanglement}

The debate introduced by EPR and Schr\"odinger's analysis regarding the definition of physical reality and correlations in QM remained silenced for almost half a century due to the deep anti-metaphysical influence within physics of Bohr's linguistic neo-Kantism, positivism and instrumentalism. As remarked by Jeffrey Bub \cite{Bub17}, ``[...] it was not until the 1980s that physicists, computer scientists, and cryptographers began to regard the non-local correlations of entangled quantum states as a new kind of non-classical resource that could be exploited, rather than an embarrassment to be explained away.''  The reason behind this shift in attitude towards {\it entanglement} is an interesting one. As Bub continues to explain: ``Most physicists attributed the puzzling features of entangled quantum states to Einstein's inappropriate `detached observer' view of physical theory, and regarded Bohr's reply to the EPR argument (Bohr, 1935) as vindicating the Copenhagen interpretation. This was unfortunate, because the study of entanglement was ignored for thirty years until John Bell's reconsideration of the EPR argument (Bell, 1964).'' Indeed, after the triumph of Bohr in the ``EPR battle'' \cite{Bohr35, EPR}, the notion of entanglement was almost completely erased by the orthodox community of physicists under the Copenhagen spell. This was until an Irish researcher called John Stewart Bell working at the Conseil Europ\'een pour la Recherche Nucl\'eaire (CERN), wrote in 1964 a paper entitled {\it On the Einstein-Podolsky-Rosen Paradox}. In this paper he was able to derive a set of statistical inequalities that restricted the correlations described by any classical local-realistic theory \cite{Bell64}. But the true breaking point for the recognition of quantum entanglement and the possibilities it implied for quantum information processing was the unwanted  result of the famous experiment performed in Orsay at the very beginning of the 1980s by Alain Aspect, Philippe Grangier and Gerard Roger \cite{AGR81}. The result was that the Bell inequality was violated by pairs of entangled spin ``particles''. As a consequence, against Einstein's and Bell's physical intuitions, the possibility for classical theories to account for such experience was completely ruled out. The experiment designed by Aspect and his team ---repeated countless times up to the present \cite{Bernien13, Hensen15}--- could not be described by any classical local-binary (realistic)\footnote{Even though the original term is ``realistic'', we prefer to add ``binary'' for reasons that will become evident in the forgoing part of the paper.} theory. The experiment was also a sign that entanglement had to be taken seriously. It was only then that quantum computation, quantum cryptography and quantum teleportation, were developed by taking {\it entanglement} as a resource \cite{Bub17}. The new notion began to rapidly populate the journals, labs, research projects and institutions all around the world. The technological era of quantum information processing had woken up from its almost half century hibernation. An hibernation, let us not forget, mainly due to the uncritical attitude of the majority of physicists who believed that Bohr had already solved everything ---and there was no reason to engage in metaphysical questions within the theory of quanta. 

With the advent of the new millennia the era of quantum information processing became rapidly one of the main centers of research and technology around the globe. It then became necessary to reach a consensus regarding the definition of quantum entanglement ---a notion which stood at the basis of all possibilities of technical analysis and development. Its definition had been clearly established by Schr\"odinger on the basis of two main notions: {\it separability} and the {\it purity} of states. The very curious situation is that while Schr\"odinger was essentially critic about the reference to ``elementary particles'' \cite{Schr50} and the introduction of ``collapses'', the contemporary research community uncritically embraced the old definition; one which was meant to show a problem rather than a solution. A good example is the explanation of entanglement provided by Mintert et al.:  
\begin{quotation}
\noindent {\small ``Composite quantum systems are systems that naturally decompose into two or more subsystems, where each subsystem itself is a proper quantum system. Referring to a decomposition as `natural' implies that it is given in an obvious fashion due to the physical situation. Most frequently, the individual susbsystems are characterized by their mutual distance that is larger than the size of a subsystem. A typical example is a string of ions, where each ion is a subsystem, and the entire string is the composite system. Formally, the Hilbert space ${\mathcal H}$ associated with a composite, or multipartite system, is given by the tensor product ${\mathcal H_1} \otimes ... \otimes {\mathcal H_N}$ of the spaces corresponding to each of the subsystems.'' \cite[p. 61]{Mintert09}}
\end{quotation}
\noindent Already this seemingly ``natural'' introduction to QM makes implicit use of an interpretation of the orthodox quantum formalism which is far from ``obvious'' or ``self evident''. First, it implies the idea that Hilbert spaces can adequately represent `physical systems'; i.e., small elementary particles such as ions. And secondly, it also implies that such particles inhabit space-time, that one can make reference to distances and that the subspaces ---which are considered as `parts' of the original Hilbert space--- describe `subsystems'.

Assuming right from the start the metaphysics of particles as a ``common sense'' {\it given} of physical representation, the story of entanglement is then told in the following manner.  In general, the Hilbert space associated with a composite system is given by the tensor product $\mathcal{H}_1\otimes\ldots\otimes \mathcal{H}_n$ of the spaces corresponding to each of the subsystems. The idea is that we should focus on a finite dimensional bipartite quantum system described by the Hilbert space $\mathcal{H}=\mathcal{H}_1\otimes \mathcal{H}_2$. After introducing {\it separability}, another essential element enters the scene, namely, the notion of {\it pure state} ---a notion intrinsically related to the definition of {\it element of physical reality}. The orthodox account of pure state rests in the following operational definition: If a quantum system is prepared in such way that one can devise a maximal test yielding with {\it certainty} (i.e., probability equal to unity) a particular outcome, then it is said that the quantum system is in a \emph{pure state}. It is then stated that the pure state of a quantum system is described by a unit vector in a Hilbert space which in Dirac's notation is denoted by $|\psi \rangle$.\footnote{As discussed in \cite{daCostadeRonde16, deRondeMassri19c} this definition is ambiguous due to the non-explicit reference to the basis in which the vector is written. It is this ambiguity which, in turn, mixes the notion of `state of a system' and `property of a system'.} Assume now that each subsystem is prepared in the following pure states $|\psi\rangle$ and $|\psi'\rangle$. The state of the composite system is then $|\psi\rangle\otimes|\psi'\rangle$. Suppose that one had access to only one of the subsystems at a time. Then, after a measurement of any local observable $A\otimes\mathbb{I}$ on the first subsystem, (where $A$ is a hermitian operator acting on $\mathcal{H}_1$, and  $\mathbb{I}$ is the identity acting on $\mathcal{H}_2$), the state of the first subsystem will be projected onto an eigenstate of $A$, but the state of the second subsystem will remain unchanged. If later on, one performs a second local measurement, now on the second subsystem, it will yield a result that is completely independent of the result of the first measurement pertaining to the first subsystem. Hence, the measurement outcomes on the two subsystems are uncorrelated between each other and only depend on their own subsystem states.

In general, depending on the basis, a pure state in $\mathcal{H}$ is given by a superposition of pure states, $|\varphi\rangle=\sum a_i|\psi\rangle_i\otimes |\psi'\rangle_i$.
For a local operator on the first subsystem, the expected value is
\[
\mbox{Tr}(A\otimes\mathbb{I}|\varphi\rangle\langle \varphi|)=
\mbox{Tr}_1(A\rho_1),\quad \rho_1:=\mbox{Tr}_2(|\varphi\rangle\langle \varphi|),
\]
where $\mbox{Tr}_1$ and $\mbox{Tr}_2$ are the partial traces over the first and second subsystem and $\rho_1$ is the reduced density matrix of the first subsystem. Then, one can conclude that the state of the first subsystem is given by $\rho_1$ and the state of the second subsystem by $\rho_2$ (where $\rho_2:=\mbox{Tr}_1(|\varphi\rangle\langle \varphi|)$). However, the state of the composite system is different from $\rho_1\otimes\rho_2$. Moreover, if one performs a local measurement on one subsystem, this leads to a state reduction of the entire system state, not only of the subsystem on which the measurement had been performed. Therefore, the probabilities for an outcome of a measurement on one subsystem are influenced by the measurements on the other distant subsystem. 
Thus, measurement results on subsystems are (classically) correlated.
\begin{definition}
States that can be written as a product of pure states are called \emph{product} or \emph{separable states}. The states which are not separable are then defined as \emph{entangled states}.
\end{definition} 
As it is explicit from its definition the notions of {\it purity} and {\it separability} play an essential role within the orthodox understanding of quantum entanglement. According to orthodoxy, that which is not separable is entangled.

\section{Is Particle Metaphysics Adequate for Quantum Theory?}

Advocated by many like a dogma, still today, the atomist picture of reality rules the mind of too many physicists ---and even philosophers of physics--- which assume it as the ``common sense'' natural representation of the physical world. As pointed out by Heisenberg \cite[p. 218]{Castellani98}: ``The strongest influence on the physics and chemistry of the past [19th] century undoubtedly came from the atomism of Democritos. This view allows an intuitive description of chemical processes on a small scale. Atoms can be compared with the mass points of Newtonian mechanics, and from this a satisfactory statistical theory of heat was developed. [...] the electron, the proton, and possibly the neutron could, it seemed, be considered as the genuine atoms, the indivisible building blocks, of matter.'' Even though it was clearly restricted by the anti-metaphysical {\it Zeitgeist} of the 20th Century, the influence of atomism did not stop with the creation of the theory of quanta. And even though QM was born from a radical departure from classical notions, this did not stop physicists from claiming that the mathematical formalism of the theory obviously referred to ``elementary particles''. In his famous paper: {\it What is an elementary particle?} Schr\"odinger strongly criticized this situation: ``Atomism in its latest form is called quantum mechanics. [...] In the present form of the theory the `atoms' are electrons, protons, photons, mesons, etc. The generic name is elementary particle, or merely particle. The term `atom' has very wisely been retained for chemical atoms, though it has become a misnomer.'' Today, more that half a century from Schr\"odinger's critical reflections, the present situation has not improved a single bit. On the contrary, it seems to have become worse since many physicists ---and even philosophers--- do not even acknowledge the fact that the atomist metaphysical picture is just one between many possible representations of reality ---and not an unescapable way to talk about Nature. 

In the context of QM, this dogmatic viewpoint has imposed the uncritical introduction of the notions of `system' and `state' within the axiomatic formulation of the theory. However, when pushed to explain what quantum systems really are, physicists ---who at least recognize the difficulties--- end up claiming that the reference to elementary particles ``is just a way of talking''. Apart from its being ``weird'' or ``quantum'', no one seems to understand what a quantum particle really is. And even though all the research in the last century points explicitly to the fact we do not understand what QM is talking about, orthodox textbooks used in Universities all around the world still teach every student that the theory of quanta talks about ``tiny elementary particles''. But particle metaphysics is not the only element within the present ``quantum omelette'' \cite{Jaynes}. The positivist presupposition imposing the addition of a ``collapse'' each time the quantum wave function is observed also plays an essential role within the many pseudoproblems presently discussed within the specialized literature. It is only with {\it pure states} that such a collapse remains unnoticed. It is only in this case ---given the choice of the correct context (or basis)--- that {\it what is} remains the same to {\it what is observed}. It is only pure states which allow us to consider observables as {\it elements of physical reality} (in the EPR sense).\footnote{The acceptance of EPR's reality criteria goes in line with the operational quantum logic approach proposed by Constantin Piron \cite{Piron76} and subsequently developed by Aerts himself \cite{Aerts81}. As Aerts and Massimiliano Sassoli de Bianchi \cite[p. 20]{AertsSassoli15} ---both students of Piron--- explain with great clarity: ``the notion of `element of reality' is exactly what was meant by Einstein, Podolsky and Rosen, in their famous 1935 article. An element of reality is a state of prediction: a property of an entity that we know is actual, in the sense that, should we decide to observe it (i.e., to test its actuality), the outcome of the observation would be certainly successful.''} Indeed, pure states guarantee the existence of an observable which is always {\it certain} (probability equal to 1) if measured. Or in other words, it is only pure states which allow an interpretation of a quantum observable in terms of an {\it actual property}; i.e., a property that will yield the answer {\it yes} when being measured. At the opposite corner, superposed states of more than one term do not describe observables which, when measured, will be obtained with certainty. The properties constituting superposed states\footnote{See \cite{deRonde18} for an explicit definition of quantum superposition.} of more than one term are referred to in the literature as {\it indeterminate}  or {\it potential} properties. Indeterminate properties might, or might not become actualized in a future instant of time and thus, cannot be considered as elements of physical reality (in the EPR sense). It is at this point that the empiricist-positivist understanding of physics as a formal scheme capable of describing observations has deeply influenced the need to introduce a `projection postulate' that would allow to transform quantum superpositions into single outcomes ---which is, as argued by orthodoxy, what we actually observe in the lab. In turn, this projection, added to the axiomatic formulation of the theory in a completely {\it ad hoc} manner, is interpreted as a real ``collapse''. In order to secure what we observe (a single `click' and not a superposition of `clicks') this invisible collaspe takes place each time an observer decides to perform a measurement. However, as remarked by Dennis Dieks \cite[p. 120]{Dieks10}: ``Collapses constitute a process of evolution that conflicts with the evolution governed by the Schr\"{o}dinger equation. And this raises the question of exactly when during the measurement process such a collapse could take place or, in other words, of when the Schr\"{o}dinger equation is suspended. This question has become very urgent in the last couple of decades, during which sophisticated experiments have clearly demonstrated that in interaction processes on the sub-microscopic, microscopic and mesoscopic scales collapses are never encountered.'' In the last decades, the experimental research seems to confirm there is nothing like a ``real collapse process'' suddenly happening when measurement takes place. Unfortunately, as Dieks \cite{Dieks18} also acknowledges: ``The evidence against collapses has not yet affected the textbook tradition, which has not questioned the status of collapses as a mechanism of evolution alongside unitary Schr\"odinger dynamics.'' 

 It is at this point that we might recall that physical research is not a process of justification of our prejudices regarding our ``common sense'' picture of reality, it is exactly the opposite. Physics begins with the humble acceptance of the unknown and continues its effort to expand our understanding and representation of reality. Within this process, critical thought is obviously essential. The addition of fictitious inadequate concepts ---such as `quantum particles' and `quantum jumps'--- to discuss the reference of the formalism of QM is not the way out of the labyrinth, it is the entrance. In fact, it is the scrambling of particle metaphysics and invisible collapses with the quantum formalism which is responsible for creating a numerous set of (pseudo)problems. The fact that the notion of `system' is inadequate to explain the formalism of QM ---something which has been exposed in an extreme manner by the superposition principle and quantum contextuality (see \cite{deRonde18, deRondeMassri16})--- or the fact that collapses have no empirical ground nor play any role within the operational application of the theory, has not stoped physicists and philosophers from repeating their mantra: QM talks about ``elementary particles'' which ``collapse'' each time we measure them. The same criticism can be applied to the notion of separability, which played an essential role both in the construction of the EPR paradox and the definition of entanglement (see for a detailed analysis: \cite{Aerts81, Aerts84b}). Such concept is also completely absent from any direct link to the orthodox formalism of QM since, as recognized by Einstein himself \cite[p. 172]{Born71}, ``quantum theory nowhere makes explicit use of this requirement.'' 

\smallskip

In the following sections we will discuss two recent approaches to QM which attempt to understand the orthodox formalism of QM without making any reference to ``tiny (separable) particles'' or strange ``collapses''. While the first goes against the addition of metaphysics and supports a neo-Bohrian linguistic understanding of Physics, the second approach attempts to present a new metaphysical representation which consistently and intuitively accounts for what QM is really taking about.

\section{The Device-Independent Approach: Beyond Particle Metaphysics?} 

Operational axiomatic approaches to QM have a long history going back to Dirac's and von Neumann's formulation of the theory during the 1930s. Later on, in the 1960s and 70s  the Geneva School commanded by Josef-Maria Jauch and Constantin Piron kept developing these ideas further in the context of quantum logic. As explained by Sonja Smets:  
\begin{quotation}
\noindent {\small ``In the language provided by the Geneva approach, a physical property is called {\it actual} if and only if the DEP's [Definite Experimental Project] which test it are certain and is {\it potential} otherwise. When a property is actual or not, depends on the state in which one considers the system to be. The Geneva approach adopts here a realistic stance towards physical properties. The underlying assumption is that the physical properties have an extension in reality, can be described and characterized by physicists and are considered to be measurable. In particular the EPR-`criterion of reality' (see \cite{EPR}) is explicitly adopted and explains why measurability is an important ingredient. Indeed, an `actual property' is closely linked to the notion of `element of reality' introduced in \cite{EPR}.'' \cite[p. 47]{Smets05}}
\end{quotation}
But while the Geneva approach adopted a realist viewpoint with respect to `properties' and `systems' (or `entities'), the more recent device-independent approach to QM ---even though continues the operational trend of thought--- remains at safe distance from realist claims. As remarked by Alexei Grinbaum:
\begin{quotation}
\noindent {\small ``Operational axiomatic approaches to quantum mechanics focus on the inputs and outputs of the observer: a `box' picture. The postulates that successfully constrain the box to behave according to the rules of quantum theory become our best candidates for fundamental principles of Nature. In a device-independent approach, such postulates are also at work: they are the only content of physical theory along with the inputs and the outputs of the parties.'' \cite{Grinbaum17}}
\end{quotation}
Together with properties, the device-independent approach drops the notion of `system' itself. The argument for doing so is quite simple: ``If a theory contains no notion of system, there is no reason to picture reality as comprised of physical entities.'' Grinbaum continues to explain: ``Systems in the device-independent approach are unnecessary not only for the purposes of interpretation, but also on the theoretical side. They cannot correspond to objective reality because they are absent from the theory. Both in the philosophy of physics and in its mathematics systems are no more a requirement.'' Of course, the idea that the formalism of QM should be regarded beyond a direct reference to an intuitive conceptual representation is not new. This idea goes obviously back to Bohr himself who denied explicitly the possibility of doing so through his doctrine of classical concepts. Bohr stressed the reference to (classical) language, and in particular, he argued that ``[...] the unambiguous interpretation of any measurement must be essentially framed in terms of classical physical theories, and we may say that in this sense the language of Newton and Maxwell will remain the language of physicists for all time.'' Furthermore, he claimed  \cite{Bohr60} that: ``Physics is to be regarded not so much as the study of something a priori given, but rather as the development of methods of ordering and surveying human experience. In this respect our task must be to account for such experience in a manner independent of individual subjective judgement and therefor objective in the sense that it can be unambiguously communicated in ordinary human language.'' Grinbaum, following the participatory realism introduced by Wheeler ---a student of Bohr---, seems to have gone a step further: ``the propositions are themselves elements of reality and [...] they do not need to refer to any entities whatsoever, whether empirical or theoretical. Device-independent models proceed on a similar view replacing Wheeler's `undecidable propositions' by an ensemble of operationally defined inputs and outputs.'' It is from this standpoint that the device-independent approach developed by Grinbaum engages in the problem of {\it reference} of the theory: if the theory is not a theory about physical systems, what would it be a theory of?
\begin{quotation}
\noindent {\small ``One  finds a tentative answer in a definition using only the strictly necessary concepts: {\it For Alice (respectively for Bob), an experiment is a process or black box to which she feeds an input x from the alphabet X and from which she receives an output a from the alphabet A. Alphabets X; Y; A; B are of  finite cardinality. [60]} On this view, physical theory is about languages: it is defined by a choice of alphabets for the inputs and the outputs and by the conditions imposed on this algebraic structure. Strings, or words in such alphabets, form a common mathematical background of device-independent approaches.'' \cite{Grinbaum17}}
\end{quotation} It is interesting to remark that there exist interesting connections between the device-independent approach presented by Grinbaum, Bohr's interpretation of QM, and the information theoretic accounts by Jeffrey Bub and Chris Fuchs (see e.g., \cite{Bub04, Bub17b, Bub19, Fuchs17, Grinbaum15}). The analysis of such relations exceed the scope of the present paper which we leave for a future work.
 


The device independent approach is critical about the notion of `system' when applied to QM. But even though Grinbaum shows that `systems' are nowhere to be found in the theory of quanta, he still takes for granted the quantum informational contemporary definition of entanglement. While the notion of `system' is dropped ---and together with it, possibly also that of separability--- the orthodox notion of entanglement ---which grounds itself on the notion of `system' and `separability'---  seems to be retained without being criticized or revisited. In fact, as we have discussed above, even though entanglement was defined in radically critical terms by both Einstein and Schr\"odinger, pointing to the difficulties and inconsistencies of the notion, its use and application within the field of quantum information has remained almost completely silent and uncritical regarding such difficulties. This fact might be related to the widespread contemporary instrumentalist understanding of physics as a discipline which only predicts measurement outcomes. Indeed, while an understanding of physics in terms of a theoretical (formal-conceptual) representation which describes ---in a particular manner--- a state of affairs, must necessarily search for the consistent introduction of adequate concepts an algorithmic understanding of physics ---which is only worried about the predictions of `clicks' in detectors--- does not need to bother about the addition of a picture. As argued by Dirac: ``it might be remarked that the main object of physical science is not the provision of pictures, but the formulation of laws governing phenomena and the application of these laws to the discovery of phenomena. If a picture exists, so much the better; but whether a picture exists of not is a matter of only secondary importance.'' Of course, nor Einstein nor Schr\"odinger agreed with this account of physics. For them, conceptual representation was a necessary condiment of a physical theory, an indispensable element even for deciding what the theory was able to observe. This divergent understanding of physics, either as an algorithm capable of predicting observations or as a theoretical representation of a state of affairs, was in fact one of the kernel points of debate between the founding fathers of QM. At the heart of this conundrum stood the role that metaphysics played within physics.

\section{The Role of Metaphysics in Physics}

Dirung the 20th Century the positivist understanding of physical theories in terms of {\it theoretical terms} and {\it empirical terms} has restricted the role of metaphysics within physics to its minimum expression. According to this trend of thought, metaphysics is only a fictional story added to an already empirically adequate theory, a created illusion required only by those metaphysically inclined researchers who wish to continue to discuss beyond what is actually observable. In this respect, even though the role of observations as empirical {\it givens} was strongly criticized within positivism itself ---firstly by Hanson, but then also by Popper, Kuhn and Feyerabend--- the (naive) empiricist understanding of theories and the problems derived from this viewpoint has remained completely unchanged. According to the standard view \cite{Redei19}, ``physics carries out precision measurements aiming at determining values of operationally defined physical quantities [and] sets up mathematical models of physical phenomena that make explicit the functional relationships among the measured quantities.'' Theories are mainly understood as sets of (mathematical) models developed in order to account for empirical observations. Within this viewpoint, Miklos Redei \cite{Redei19} has characterized the role played by mathematics within physics in terms of what he calls the `supermarket picture': ``mathematics is like a supermarket and physics is a customer. That is to say, it is tacitly assumed that when a physicist needs a mathematical concept, a mathematical structure, or any mathematical tool to formulate the mathematical model of a physical phenomenon, then (s)he just goes to the mathematics-supermarket, looks at the shelves [and] takes off the product needed.'' The `supermarket picture' discussed by Redei is also perfect for characterizing the standard {\it praxis} of physicists and philosophers of physics when dealing with the `interpretation' of theories. When a physicist needs a (metaphysical) concept in order to build up an `interpretation', then she just goes to the `concept-supermarket', looks in the shelves and takes the needed notions. A few quantum particles, a bit of measurement interaction, some agents, maybe some worlds or minds or flashes; she puts them all together in a paper and {\it voil\`a}... a new interpretation has been cooked! What is important to remark that according to this positivist scheme, even though the `mathematics supermarket' is essential for the survival of the working physicist, the `concept-supermarket' is clearly not. The latter is only used by fancy customers who are interested in creating a (metaphysical) `picture' of what the world is like would  the theory be true \cite{VF09}. Of course, more down to earth customers who already possess an empirically adequate theory and actually do care about their money\footnote{As argued by Fuchs in \cite{Fuchs02}: ``The issue remains, when will we ever stop burdening the taxpayer with conferences devoted to the quantum foundations?''} do not need to venture into the suburbs in order to find these expensive metaphysical chains full of weird fancy concepts. 

\smallskip

There is however also a completely different ---positive--- understanding of the role and meaning of the term `metaphysics' within physical theories. According to this understanding, shared by many of the founding fathers of QM, `metaphysics' is essentially a relational categorical system of concepts. Concepts do not make reference to `things'. Concepts are always related to other concepts creating a net which allows us to conceive and represent a specific type of experience. There is no way to discuss about experience without presuposing a conceptual representation. A very good example of how a metaphysical system works is provided by Aristotle's hylomorphic scheme in which the notion of {\it entity} was characterized in terms of two modes of existence: the potential and the actual. In particular, the actual mode of existence ---which is the only one that survived after Newton's actualist metaphysical choice \cite{deRonde17}--- characterized entities in terms of three logical and ontological principles: existence, non-contradiction and identity (see for a detailed analysis \cite{deRondeBontems11}). These principles, which are not to be found in the empirical world, allow us to predicate existence of `something', to claim that the existent has non-contradictory properties, and furthermore that this non-contradictory existent remains identical to itself thorough time. Acting together ---and only together--- these principles allow us to think about an object of experience; they are thus the conditions of possibility to access experience in a systematic relational manner. In this respect, we might recall Einstein's remark to Heisenberg that: ``It is only the theory which decides what can be observed.'' This, in fact, was according to Einstein, the really significant philosophical achievement of Kant:  
\begin{quotation}
\noindent {\small``From Hume Kant had learned that there are concepts (as, for example, that of causal connection), which play a dominating role in our thinking, and which, nevertheless, can not be deduced by means of a logical process from the empirically given (a fact which several empiricists recognize, it is true, but seem always again to forget). What justifies the use of such concepts? Suppose he had replied in this sense: Thinking is necessary in order to understand the empirically given, {\it and concepts and `categories' are necessary as indispensable elements of thinking.}'' \cite[p. 678]{Einstein65} (emphasis in the original)}
\end{quotation} 
Any scientific discourse must always presuppose a conceptual representation of what is meant by a `state of affairs'. This is not ---at least for the metaphysician--- something ``self evident'' nor part of the ``common sense'' of the layman but the very precondition for understanding phenomena in a scientific manner. It is the recognition of the need of metaphysical representation, through the systematic careful creation of a net of concepts, which allows science to be critical about its own foundation. This means that willingly or not, we physicists, are always producing our praxis {\it within} a specific representation, we always observe from {\it within} a theory. From this viewpoint, representation is always first, experience and perception are necessarily second. Paraphrasing Wittgenstein's famous remark regarding language, the physical representation we inhabit presents the limits of the physical world we understand.\footnote{Let us notice, firstly, that ``physical'' should not be understood as a {\it given} ``material reality'', but rather as a procedure for representing reality in theoretical ---both formal and conceptual--- terms. And secondly, that the relation between such physical representation and reality is not something ``self evident''. The naive realist account according to which representation ``discovers'' an already ``fixed'' reality is not the only possibility that can be considered. A one-to-one correspondence relation between theory and reality is a very naive solution to the deep problem of relating theory and {\it physis.}} This marks a point of departure with respect to naive empiricism and positivism. A point that was also stressed by Einstein: 
\begin{quotation}
\noindent {\small ``I dislike the basic positivistic attitude, which from my point of view is untenable, and which seems to me to come to the same thing as Berkeley's principle, {\it esse est percipi.} `Being' is always something which is mentally constructed by us, that is, something which we freely posit (in the logical sense). The justification of such constructs does not lie in their derivation from what is given by the senses. Such a type of derivation (in the sense of logical deducibility) is nowhere to be had, not even in the domain of pre-scientific thinking. The justification of the constructs, which represent `reality' for us, lies alone in their quality of making intelligible what is sensorily given.'' \cite[p. 669]{Einstein65}}
\end{quotation} 

The essential role played by metaphysics within physics is the creation of adequate conceptual nets each of which allow us to capture a specific field of experience. As remarked by Heisenberg \cite[p. 264]{Heis73}: ``The history of physics is not only a sequence of experimental discoveries and observations, followed by their mathematical description; {\it it is also a history of concepts.}  For an understanding of the phenomena the first condition is the introduction of adequate concepts. {\it Only with the help of correct concepts can we really know what has been observed.}'' But the creation of new physical concepts is not an easy task, it is a difficult process which requires breaking the chains of old ``common sense'' representations: 
\begin{quotation}
\noindent {\small ``Concepts that have proven useful in ordering things easily achieve such an authority over us that we forget their earthly origins and accept them as unalterable givens. Thus they come to be stamped as `necessities of thought,' `a priori givens,' etc. The path of scientific advance is often made impossible for a long time  through such errors. For that reason, it is by no means an idle game if we become practiced in analyzing the long common place concepts and exhibiting those circumstances upon which their justification and usefulness depend, how they have grown up, individually, out of the givens of experience. By this means, their all-too-great authority will be broken. They will be removed if they cannot be properly legitimated, corrected if their correlation with given things be far too superfluous, replaced by others if a new system can be established that we prefer for whatever reason.'' \cite[p. 102]{Einstein16}}
\end{quotation}

Of course, we do not believe that the structural relationship between the formal and the conceptual levels of a theory is developed in a linear or straightforward manner. On the contrary, it is in general an entangled interrelated process that goes back and forth between the creation of new mathematics, conceptual schemes and even technical developments. One level helps the other. Unfortunately, the orthodox attempt rather than discussing this relation and developing new adequate concepts has been focused ---following Bohr's correspondence principle--- in ``bridging the gap'' between the mathematical formalism and our manifest (classical) image of the world \cite{Dorato15}. Instead, taking the mathematical formalism as a standpoint and advancing beyond Bohr's prohibitions, one can also attempt to develo a new adequate non-classical conceptual framework. This requires a careful analysis of the conditions implied by the mathematical formalism of the theory. Following such type of analysis,\footnote{The fact that this notion is not adequate in order to interpret the quantum formalism is explicit from the Kochen-Specker theorem \cite{deRonde17, deRondeMassri16, deRondeMassri18}, the existence of quantum superpositions \cite{deRonde18, deRondeMassri19} and the non-separability theorem \cite{Aerts81, Aerts84b}.} as we argued in \cite{deRondeBontems11}, the notion of {\it entity} (`system' or `object') even though is essential for classical mechanics, becomes in the context of quantum theory an {\it epistemological obstruction}, an element retained from a purely dogmatic metaphysical standpoint. 

If we accept that there must exist an adequate structural relationship between mathematical formalisms and conceptual frameworks it is not difficult to see that the quantum formalism cannot represent `separable systems'. While the equation of motion in classical mechanics can be expressed in ${\cal R}^3$ allowing an interpretation in 3-dimensional Euclidean space, QM works in a configuration space. The difference is essential when attempting to consider existents within space. Classically, if we add two systems, the properties are summed. Given two systems with a number of properties, $R$ and $R'$, respectively; their joint consideration is just the sum of the properties of each system, namely, $R+R$. A paradigmatic example is the completely inelastic crash of two systems. While before the crash the two particles are separated and their mass are $m$ and $m'$, and their velocities are $v$ and $v'$, respectively; after the crash they become a (non-separable) single system of mass $m+m'$ with a common velocity $v_f$. The essential property characterizing the two systems ---namely, their mass--- becomes nothing else than the sum of masses. But, as we know, there is an essential difference when considering the addition of `systems' (vector spaces) in QM, $\mathcal{H}=\mathcal{H}_1\otimes \mathcal{H}_2$. If we take two rays which intersect each other, in terms of classical set theory, the addition of the rays is just the two rays; however, in terms of vector spaces the addition of two rays (now considered as subspaces, $\mathcal{H}_1$ and $\mathcal{H}_2$) is more than just their sum. In fact, $\mathcal{H}$ is the whole plane {\it generated} by the two rays (see also the analysis provided by Rob Griffiths in \cite[Sect. 2]{Griffiths02}). In QM, the new possibilities considered by the addition of systems are not just the sum of the previous subsystems, they are much more.\footnote{In this respect, the logos approach provides an intuitive understanding of what is going on in terms of the capabilities of an apparatus: adding two apparatuses allows many more possibilities than just the reductionistic sum of their previous possibilities. Projection operators are not properties, but possibilities of action related to degrees of freedom ---which is what {\it configuration space} is actually about.} An excellent example of this problematic situation within orthodox QM is discussed by Rob Clifton in \cite{Clifton95, Clifton96} where he analyzes from a classical perspective the inconsistency present in QM when attempting to provide a valuation of the properties pertaining to a `system' and its `subsystems'.\footnote{Clifton developed an example in which the violations of {\it Property Composition} and {\it Property Decomposition} seem to show implications which seem at least incompatible with the everyday description of reality. In the example Clifton takes a Boeing 747 which has a possibly wrapped left-hand wing: $\alpha$ is the left-hand wing and $\alpha \beta$ is the airplane as a whole. $Q_{\alpha}$ represents the property of being wrapped and $Q_{\alpha \otimes I_\beta}$  represents the plane property of the left wing being wrapped. In such an example a violation of {\it Property Composition} ($[Q_{\alpha }] = 1, [Q_{\alpha \otimes I_\beta}] \neq 1$) leads, according to Clifton \cite[p. 385]{Clifton96}, to the following situation: ``a pilot could still be confident flying in the 747despite the fault in the left hand wing''. If, on the other hand, {\it Property Decomposition} fails ($[Q_{\alpha \otimes I_\beta}] = 1, [Q_{\alpha }] \neq 1$) the implication reads ``no one would fly in the 747; but, then again, a mechanic would be hard-pressed to locate any flow in its left-hand wing''. The situation gets even stranger when the pilot notices that the plane as a whole has the property $[Q_{\alpha \otimes I_\beta}] = 1$ and concludes (incorrectly) following {\it Property Decomposition} that the left-hand wing is wrapped, that is, that  $[Q_{\alpha }] = 1$. The mechanic is then sent to fix the left hand-side wing but according to Clifton cannot locate the flaw because the wing does not possess the property $Q_{\alpha}$} While the logic of classical mechanics follows that of Boolean sets, the logic of quantum mechanics is non-distributive and the {\it joint} is clearly non-classical. In the context of quantum logic, Diederik Aerts has even derived a non-separability theorem which shows that quantum systems are essentially non-separable \cite{Aerts81, Aerts84b}. All these different results might be regarded as what Wolfgang Pauli would call ``road signs'', all of which point in the same direction, namely, the necessity to go beyond the classical notion of separability imposed by particle metaphysics. As already remarked by Grinbaum ``If a theory contains no notion of system, there is no reason to picture reality as comprised of physical entities.'' Contrary to the dogma professed by contemporary atomists there is no reason to believe that the only way to picture reality in physics is through the notion of `system'.

\section{Intensive Relational Metaphysics and Entanglement} 

Today, orthodoxy assumes that physics describes `systems' in an algorithmic fashion which allows us to predict observable measurement outcomes. This understanding of physics restricted by the classical paradigm ---mainly due to Bohr's philosophy of physics supplemented by 20th Century positivism and, later on, also instrumentalism--- has blocked the possibility to advance in the development of a new conceptual scheme for understanding QM. The danish physicist was explicit regarding this point and argued repeatedly that: ``it would be a misconception to believe that the difficulties of the atomic theory may be evaded by eventually replacing the concepts of classical physics by new conceptual forms.'' Exactly this type of warning, is what David Deutsch \cite{Deutsch04} has rightly characterized as ``bad philosophy'', namely, ``[a] philosophy that is not merely false, but actively prevents the growth of other knowledge.''\footnote{David Deutsch continues his explanation with a direct attack to the Bohrian philosophy: ``The physicist Niels Bohr (another of the pioneers of quantum theory) then developed an `interpretation' of the theory which later became known as the `Copenhagen interpretation'. It said that quantum theory, including the rule of thumb, was a complete description of reality. Bohr excused the various contradictions and gaps by using a combination of instrumentalism and studied ambiguity. He denied the `possibility of speaking of phenomena as existing objectively' ---but said that only the outcomes of observations should count as phenomena. He also said that, although observation has no access to `the real essence of phenomena', it does reveal relationships between them, and that, in addition, quantum theory blurs the distinction between observer and observed. As for what would happen if one observer performed a quantum-level observation on another, he avoided the issue ---which became known as the `paradox of Wigner's friend', after the physicist Eugene Wigner.''} Breaking the Bohrian law, the logos approach to QM attempts to develop a new metaphysical scheme with specially suited non-classical concepts that are able to explain in an intuitive manner what QM is really talking about. In fact, by carefully studying the orthodox quantum formalism it is possible to derive an important set of consequences which have been always there, at plain sight. 

To take the formalism seriously means for us to seek for an objective set of concepts which are grounded on the mathematical structure of the formalism itself. In particular, as we have argued elsewhere \cite{deRondeMassri16}, the key to understand the objective aspect of the mathematical formalism of QM is not something related to observations, it is exposed in the invariant mathematical structure of the theory. As Max Born himself reflected: \cite{Born53}: ``the idea of invariant is the clue to a rational concept of reality, not only in physics but in every aspect of the world.'' In physics, invariants are quantities having the same value for any reference frame. The transformations that allow us to consider the physical magnitudes from different frames of reference have the property of forming a group. It is this feature which allows us to determine what can be considered {\it the same} according to a mathematical formalism. In the case of classical mechanics invariance is provided via the Galilei transformations while in relativity theory we find the Lorentz transformations. In QM the invariance of the theory is exposed by no other than Born's famous rule.

\smallskip
\smallskip

\noindent {\it
{\bf Born Rule:} Given a vector $\Psi$ in a Hilbert space, the following rule allows us to predict the average value of (any) observable $P$. 
$$\langle \Psi| P | \Psi \rangle = \langle P \rangle$$
This prediction is independent of the choice of any particular basis.}

\smallskip
\smallskip

\noindent This rule, which provides the invariant structure of the theory, points implicitly to the way in which physical reality should be conceived according to the theory of quanta. Taking distance from the famous Bohrian prohibition to consider physical reality beyond the theories of Newton and Maxwell, we have proposed the following extended definition of what can be naturally considered ---by simply taking into account the mathematical invariance of the Hilbert formalism--- as a generalized element of (quantum) physical reality (see \cite{deRonde16}).

\smallskip
\smallskip

\noindent {\it {\bf Generalized Element of Physical Reality:} If we can predict in any way (i.e., both probabilistically or with certainty) the value of a physical quantity, then there exists an element of reality corresponding to that quantity.}

\smallskip
\smallskip

\noindent This redefinition implies a deep reconfiguration of the way in with the quantum formalism must be addressed, the type of predictions it provides and even the way in which data must be analyzed (see for a detailed discussion \cite{deRondeFreytesSergioli19}). It also allows us to understand Born's probabilistic rule in a new light; not as providing information about a (subjectively observed) measurement result, but instead, as providing (objective) information of a theoretically described (potential) state of affairs \cite{deRonde16}. Objective probability does not mean that particles behave in an intrinsically random manner. Objective probability means that probability characterizes a feature of the conceptual representation accurately and independently of any subjective choice or observation ---i.e., in invariant terms. This account of probability allows us to restore a representation in which the state of affairs is detached from the observer's choices to measure (or not) a particular property ---just like Einstein's account of physics in terms of {\it detached observers} requested. Consequently, the Born rule always provides complete knowledge of the quantum mechanically described state of affairs; in cases where the probability is equal to 1 and also in cases in which probability is different to 1. Any vector or matrix, independently of the context (or basis), provides {\it maximal knowledge} of the represented (quantum) state of affairs. Since there is no essential mathematical distinction between any matrix (of any rank), both pure states and mixtures have to be equally considered; none of them is ``less real'' or ``less well defined'' than the others. Thus, it is not necessary at all to distinguish between pure states and mixed states.\footnote{This point has been already addressed by David Mermin in \cite[Sect. VII]{Mermin98}.} In turn, through the strict application of the Born rule in order to define {\it intensive valuations} we have also been capable to derive a non-contextuality intensive theorem which bypasses the Kochen-Specker theorem in a natural manner and allows us to restore a global objective reference to all projection operators without any inconsistency \cite{deRondeMassri18}. 

At safe distance from many approaches which assume a classical metaphysical standpoint when analyzing QM ---introducing implicitly or explicitly classical notions within the theory---, the logos approach has been devised as an account of QM which stays close to the quantum formalism in the most strict manner. This implies for us, a suspicious attitude towards the (classical) notions of `system', `state' and `property'. Taking their place, we have created new (non-classical) concepts which attempt to satisfy the features of the quantum formalism ---and not the other way around. According to the logos approach, QM talks about a potential realm which is independent of actuality. There is never a ``collapse'' from a quantum superposed state to a measurement outcome, simply because physics does not describe the observations of agents. Physics represents in a formal-conceptual manner states of affairs and their evolution. Following this understanding, we have argued that QM talks about a potential state of affairs constituted by immanent powers with definite potentia. From this standpoint, we have shown how through the aid of these newly introduced notions we are able to explain the distance between the objective representation provided by the theory and the subjective measurements taking place {\it hic et nunc} in a lab  \cite{deRondeMassri18} ---dissolving in this way the infamous measurement riddle. Forced by the need to replace particle metaphysics and collapses with a new adequate metaphysical scheme, in \cite{deRondeMassri19b} we have also derived a new objective definition of entanglement in terms of the potential coding of intensive and effective relations. 

\medskip

\noindent {\it
{\sc Effective Relations:} The relations determined by a difference of possible actual effectuations. Effective relations discuss the possibility of an actualist definite coding. It involves the path from intensive relations to definite correlated (or anti-correlated) outcomes. They are determined by a binary valuation of the whole graph in which only one node is considered as true, while the rest are considered as false.}

\medskip

\noindent {\it
{\sc Intensive Relations:} The relations determined by the intensity of different powers. Intensive relations imply the possibility of a potential intensive coding. They are determined by the correlation of intensive valuations.}

\medskip 

\noindent These relations provide an intuitive grasp of what can be done in a lab and what type of relations are at play. The following definitions provide a new account of entanglement which rests on the analysis of relational intensive and effective correlations.
\begin{definition}[Quantum Entanglement]
Given $\Psi_1$ and $\Psi_2$ two PSAs, if $\Psi_1$ and $\Psi_2$ are related intensively and effectively we say there exists \emph{quantum entanglement} between $\Psi_1$ and $\Psi_2$.
\end{definition}
According to this definition entanglement relates to the potential coding of intensive and effective relations between two distant measuring set-ups. We also have the possibility to provide an intuitive non-spatial definition of separability which relates to the lack of correlations between two distant screens.
\begin{definition}[Relational Separability]
Given $\Psi_1$ and $\Psi_2$ two PSAs, if $\Psi_1$ and $\Psi_2$ are not related intensively nor effectively we say there is \emph{relational separability} between $\Psi_1$ and $\Psi_2$.
\end{definition}
It is interesting to notice that our definitions of potential coding in terms of intensive and effective relations allows us to address a third possibility which considers the cases in which there are only intensive relations involved but not effective ones. 

\begin{definition}[Intensive Correlation]
Given $\Psi_1$ and $\Psi_2$ two PSAs, if $\Psi_1$ and $\Psi_2$ are related intensively but not effectively we say there exists an \emph{intensive relation} between $\Psi_1$ and $\Psi_2$.
\end{definition}

This new approach shows how metaphysical considerations are essential for the analysis of operational data. In fact, the analysis of intensive relations has been completely bypassed within the orthodox definition of entanglement, focused on the collapse of invisible particles characterized in terms of properties described in binary terms. From this viewpoint, what needs to be stressed is that since the notion of entanglement is grounded explicitly on both `particle metaphysics' and the `collapse' of the quantum wave function, the rejection of these elements ---present within the orthodox axiomatic Dirac-von Neumann formulation of the theory--- also implies the rejection of the present definition of quantum entanglement. Our redefinition of the notion of entanglement beyond classical notions and {\it ad hoc} rules hopes to open the debate about such a possibility.


\section{Conclusion} 

In this paper we have provided arguments against the orthodox definition of quantum entanglement as grounded on `particle metaphysics' and the existence of unobserved ``collapses'' added to the axiomatic formulation of the theory in a completely {\it ad hoc} manner. We have discussed and analyzed two different approaches which attempt to go beyond the notion of `system' in QM. While the first device-independent approach retains the orthodox definition of entanglement, the logos approach presents a new definition which requires the consideration of intensive relations.

\section*{Acknowledgements} 

We want to thank an anonymous referee for his comments and remarks which have allowed us to change the manuscript substantially. C. de Ronde would like to thank Don Howard for historical references. We would also like to thank Dirk Aerts and Massimiliano Sassoli de Bianchi for related discussions. This work was partially supported by the following grants: FWO project G.0405.08 and FWO-research community W0.030.06. CONICET RES. 4541-12 and the Project PIO-CONICET-UNAJ (15520150100008CO) ``Quantum Superpositions in Quantum Information Processing''.

\end{document}